\documentclass[prd, reprint, amsmath,amssymb,nofootinbib,floatfix, aps, longbibliography,  superscriptaddress, preprintnumbers]{revtex4-2}
\usepackage{graphicx}% Include figure files
\usepackage{dcolumn}% Align table columns on decimal point
\usepackage{bm}% bold math
\usepackage{graphicx}
\usepackage{dcolumn}
\usepackage{bbold}
\usepackage{mathtools}
\usepackage{siunitx}
\usepackage[colorlinks=true,urlcolor=rossos,anchorcolor=black,citecolor=mygreen,filecolor=black,linkcolor=black,menucolor=blue,linktocpage=true]{hyperref} % should be commented out if the tex file will be compiled with latex in arXiv!!! (pdflatex is fine)
\usepackage{booktabs}
\usepackage{colortbl}
\usepackage{collcell}
\usepackage{enumerate}
\usepackage{ulem}
\usepackage{float}
\usepackage{verbatim}
\usepackage{multirow}
\usepackage{xparse}
\usepackage{tabularx}
\usepackage{relsize}
\usepackage{xcolor}
\usepackage{siunitx} %prints error unknown parameter "physics" (seems harmless)

\DeclareSIUnit\clight{\text{\ensuremath{c}}}
\DeclareSIUnit\tevm{\TeV\per\clight\squared}
\DeclareSIUnit\events{events}
\DeclareSIUnit\years{y}
\DeclareSIUnit\tonne{t}
\DeclareSIUnit\tonneyears{\tonne\years}
\DeclareSIUnit\cmtwo{\cm\squared}
\DeclareSIUnit\kev{\kilo\eV} 
\DeclareSIUnit\mev{\mega\eV} 
\DeclareSIUnit{\GeV}{\giga\eV}
\DeclareSIUnit{\TeV}{\tera\eV}
\DeclareSIUnit{\kevm}{\keV\per\clight\squared}
\DeclareSIUnit{\mevm}{\MeV} %we use c=1 for DM mass
\DeclareSIUnit{\gevm}{\GeV}
\DeclareSIUnit{\tevm}{\TeV}
\DeclareSIUnit\kevnr{\si{\kev} {}_\mathrm{NR}} 
\DeclareSIUnit\kever{\si{\kev} {}_\mathrm{ER}} 
\DeclareSIUnit\PE{\mathrm{PE}}
\DeclareSIUnit\bar{bar}

\newcommand{\virg}[1]{``#1''}

\definecolor{rossos}{cmyk}{0,1,1,0.55}

\definecolor{purple}{rgb}{0.7,0,1}
\definecolor{teal}{rgb}{0,0.5,0.5}

\definecolor{mygreen}{rgb}{0.2, 0.64, 0.48}
\definecolor{mygray}{gray}{.95}

\def\DM{\mathsmaller{\mathrm{DM}}}
\def\BH{\mathsmaller{\mathrm{BH}}}
\def\GS{\mathsmaller{\mathrm{GS}}}
\def\Baryon{\mathsmaller{\mathrm{B}}}

\def\max{{\mathrm{max}}}
\def\min{{\mathrm{min}}}
\def\tot{{\mathrm{tot}}}
\def\txsname{{TXS~0506+056}}

\begin{document}

\preprint{APS/123-QED}

\title{Setting limits on blazar-boosted dark matter\\
with xenon-based detectors}% Force line breaks with \\

% \author{Elena Aprile}
% \affiliation{Physics Department, Columbia University, New York, NY 10027, USA}

\author{Erin Barillier}
\affiliation{Physics Institute, University of Zurich, Winterthurerstrasse 190/Building 36, 8057 Zurich, Switzerland}
 
\author{Laura Manenti}
\homepage{laura.manenti@sydney.edu.au}
\affiliation{The School of Physics, The University of Sydney, Caperdown, Sydney, Australia}

\author{Knut Mor\aa}
\affiliation{Physics Institute, University of Zurich, Winterthurerstrasse 190/Building 36, 8057 Zurich, Switzerland}
 
\author{Paolo Padovani}
%\email{ppadovan@eso.org}
\affiliation{European Southern Observatory, Karl-Schwarzschild-Str. 
2, D-85748 Garching bei M\"unchen, Germany}
\affiliation{Center for Astrophysics and Space Science (CASS), New York University Abu Dhabi, PO Box 129188 Abu Dhabi, United Arab Emirates}

\author{Isaac Sarnoff }
\homepage{sarnoff@nyu.edu}

\affiliation{Center for Astrophysics and Space Science (CASS), New York University Abu Dhabi, PO Box 129188 Abu Dhabi, United Arab Emirates}

\author{Yongheng Xu}
\affiliation{Department of Physics, University of Oslo, N-0316 Oslo, Norway}
\affiliation{Department of Physics and Astronomy, University of California, 475 Porto Plaza, California 90025, USA}

\author{Bj\"{o}rn Penning}
\affiliation{Physics Institute, University of Zurich, Winterthurerstrasse 190/Building 36, 8057 Zurich, Switzerland}

\author{Francesco Arneodo}
\affiliation{Center for Astrophysics and Space Science (CASS), New York University Abu Dhabi, PO Box 129188 Abu Dhabi, United Arab Emirates}

\date{\today}% It is always \today, today,

\begin{abstract}

Dual-phase xenon time projection chambers achieve optimal sensitivity for dark matter in the 10 to \SI{1000}{\GeV c^{-2}} mass range, but sub-GeV dark matter (DM) particles lack sufficient energy to produce nuclear recoils above detection thresholds in these detectors.  
Blazar-boosted dark matter offers a way to overcome this limitation. Relativistic jets in active galactic nuclei can accelerate light dark matter in their host-galaxy halos to energies that can leave detectable nuclear recoil signals in xenon-based detectors on Earth.
We present the first blazar-boosted dark matter search that incorporates detector response modeling, using public data from XENON1T and LZ for the blazar \txsname. We model dark matter-proton scattering in the jet environment, covering the full process from jet acceleration through to detector response.  We explore how the host galaxy dark matter density profile impacts our analysis. 
We set model-dependent exclusion regions on the dark-matter-nucleon scattering cross section for $m_\chi \simeq \SI{1}{\MeV}$ dark matter, between $\SI{5.8e-31}{\square\centi\metre}$ and $\SI{6.3e-29}{\square\centi\metre}$ using XENON1T data, and between $\SI{9.9e-32}{\square\centi\metre}$ and $\SI{2.5e-28}{\square\centi\metre}$ from LZ effective field theory (EFT) dark matter searches.
Our results show that astrophysical uncertainties, especially those in the dark-matter distribution near the supermassive black hole, are the main limitation of this search rather than detector effects. 
The limits are therefore model-dependent and should be seen as exploratory. 
They highlight both the potential and the present uncertainties of blazar-boosted dark matter as a probe of light dark matter.

%\begin{description}
%\end{description}
\end{abstract}

\maketitle

%\tableofcontents

%**********************************************
%\section{\label{sec:level1}First-level heading:\protect\\ 
%The line break was forced \lowercase{via} %\textbackslash\textbackslash}

%***************************************
\section{\label{sec:intro}Introduction}
%***************************************
The dark matter (DM) problem emerged from astronomical and cosmological observations that reveal a discrepancy between the observed gravitational effects and the visible matter in the Universe. This discrepancy requires either additional undetected mass or modifications to our current understanding of gravity. One proposed explanation involves dark matter particles, a form of nonbaryonic matter~\cite{Bertone2005Jan}.

Among direct detection experiments searching for nonrelativistic DM particles from the galactic halo, liquid xenon-based detectors currently achieve the highest sensitivity. These detectors operate by measuring the energy deposited when a DM particle elastically scatters off a xenon nucleus, causing the nucleus to recoil. This nuclear recoil produces scintillation light and ionization electrons, which are detected to reconstruct both the position and energy of the particle interaction.
The leading experiments in this category achieve optimal detection sensitivity in the \qtyrange{10}{1000}{\gevm} weakly interacting massive particle dark matter (WIMP DM) mass range\footnote{Throughout this paper we use natural units where $c = 1$, so that masses are expressed in units of energy (e.g., GeV) rather than energy/$c^2$ (e.g., GeV/$c^2$).}. Below this range, WIMP DM particles in the local halo transfer insufficient energy to produce nuclear recoil signals above the $\sim$$\SI{3}{\keV}$ energy thresholds of these detectors, limiting the ability to search for sub-\si{\gevm}~DM. 
However, if the DM were accelerated to energies above those predicted by the standard galactic rotational velocity distribution, the same detectors may be sensitive also to lighter DM. 
Several mechanisms can provide this acceleration (``boost''), including solar reflection~\cite{an2021solar}, cosmic ray boosting~\cite{Bringmann:2018cvk, cui2022search, abe2023search, maity2024cosmic, LZ:2025iaw}, and blazar boosting~\cite{Wang:2021jic,Granelli:2022ysi}. 

Active galactic nuclei (AGNs) are the luminous central regions of galaxies powered by supermassive black holes. Blazars represent a special subclass of AGNs where jets are oriented toward Earth, making them appear exceptionally bright~\cite{Urry1995Sep,AGNReview}.

If DM interacts with ordinary matter, the jet may accelerate particles of the host galaxy DM halo.
This process can generate a flux of DM particles along the jet's axis. 
These blazar-boosted dark matter (BBDM) particles can reach Earth with energies orders of magnitude higher than typical local halo DM particles.  

This paper is organized as follows: in Section~\ref{sec:model}, we present the selection criteria for the blazar as a boosted-DM source and model both its jet physics and the DM profile of the host galaxy. In Sec.~\ref{sec:source_to_detector}, we calculate the BBDM flux, including the DM upscattering mechanism within the jet and Earth attenuation effects. 
Section~\ref{sec:detector} describes the nuclear recoil interactions in liquid xenon detectors, essentially computing the recoil spectra. In Sec.~\ref{sec:methods_limits}, we detail our approach for setting constraints on blazar-boosted dark matter using publicly available data from the XENON1T and LZ detectors. Section~\ref{sec:conclusion} presents our conclusions.

%********************************************************
\section{\label{sec:model}Blazar and dark matter at source}
%********************************************************

%********************************************
\subsection{\label{sec:TXS} Selection of blazar-boosting source} 
%********************************************
%Justification for source selection; known parameters and uncertainties
In 2013, the IceCube Neutrino Observatory~\cite{icecubewebsite} 
detected the first high-energy astrophysical 
neutrino with energies exceeding \SI{1}{PeV}~\cite{IceCube_2013}. Since 
then, it has compiled a growing list of such events (see Ref.~\cite{Icecat1} and references 
therein). 
Yet, so far, only the Galactic plane~\cite{icecube2023observation} (at 4.5 $\sigma$), and two astronomical objects have been associated with high-energy neutrinos detected by IceCube with a significance greater than $\sim 3\,\sigma$: the blazar \txsname~ at $z=0.3365$ (at the 3--3.5\,$\sigma$ level~ \cite{IceCube1, IceCube2}, although more recent observations are consistent with a [not significantly] higher p-value~\cite{IceCube_ICRC_2024})
and the local Seyfert 2 galaxy NGC\,1068 at $z = 0.004$ (at the $4.2\,\sigma$ level~
\cite{IceCube:2022der, padovani2024high}). Many more 
lower-level associations between blazars and IceCube neutrinos have been suggested 
in the literature (e.g., \cite{GP_2021,IceCube_2021,Sahakyan_2023} and references 
therein), which is consistent with a scenario where the contribution of the whole 
blazar class to the IceCube signal is not dominant but still relevant (e.g., Ref.~\cite{Bellenghi_2023}).

In this paper, we focus on \txsname, the only blazar associated with high-energy neutrinos detected by IceCube, which provides robust evidence of high-energy protons in its jet. Additionally, \txsname~ is among the brightest and best-modeled blazars, providing reliable observational constraints for our calculations. 

%Some other blazars have been tentatively associated with IceCube neutrinos, albeit with relatively low significance ($\lesssim 3\, \sigma$). For example, PKS 0735+178, at $z=0.65$, has been fitted with lepto-hadronic models which give a range of $L_p \sim 3 \times 10^{47} - 10^{50}$ erg s$^{-1}$ \cite{PKS0735}, which translates into an 
%$L_p/d_L^2$ value up to 3 times larger than the one of \txsname. %Similarly, PKS 1502+106, at $z=1.8385$, has been inferred to have a very large range of proton luminosities. Taking a typical value of $L_p \sim 10^{50}$ erg s$^{-1}$ \cite{Oikonomou_2021} implies an $L_p/d_L^2$ value a factor of 4 smaller than the one of \txsname, mainly because of the very large redshift of this source. 

If more reliable neutrino associations are made for other well-modeled blazars, this type of analysis could be extended to multiple blazars, potentially increasing the total BBDM flux substantially and reducing the reliance on modeling a single blazar.

We consider only proton-WIMP interactions in the jet as the boosting mechanism for two reasons. First, since we are investigating nuclear recoils in liquid xenon, we adopt the same assumption used by xenon-based experiments (i.e., that WIMPs interact with nucleons). We are therefore adding no new assumptions by stating that WIMP-nuclear interactions also occur in the blazar. Second, as the hadronic luminosity in blazar jets exceeds the leptonic luminosity, we assume proton interactions are the dominant contribution to the blazar-boosted dark matter signal. 
Additionally, we assume that the proton-WIMP cross section remains constant with respect to the WIMP kinetic energy.

Since light travel time from this blazar is 3.75\,Gyr~\cite{1802.01939}, any particles it emits must travel at least this long to reach Earth. Given that the Universe is 13.8\,Gyr old, and that galaxies might form 0.2\,Gyr after the big bang~\cite[e.g.,][]{Maiolino2024Mar}, the blazar's maximum possible age is about 9.8 Gyr ($13.8-3.75-0.2=9.8$\,Gyr).
For BBDM particles to be detectable on Earth, they must complete their journey within this 9.8\,Gyr time frame. This constraint establishes a minimum velocity requirement for these particles, corresponding to a Lorentz factor of at least 1.1. Since this is very close to $\gamma = 1$ (i.e., the nonrelativistic limit), this condition effectively places no meaningful constraint on the particles' velocities. 

In this analysis, we treat the blazar as a steady-state source. Although blazars exhibit flaring episodes lasting from hours to months, these short-duration events become completely smeared out over the \SI{3.75}{\giga\years} light travel time to Earth. Small variations in boost factor during a flare correspond to tiny velocity differences ($\sim 10^{-9}$) for the boosted dark matter particles. Over the long journey, these small velocity differences accumulate into arrival time spreads of years. Consequently, the arrival times of particles from a flare that originally lasted only hours to months become spread over many years at Earth, effectively erasing the original time structure. Therefore, we observe only the time-averaged luminosity of the blazar, justifying our steady-state approximation.

%In this analysis, we treat the blazar as a steady-state source. Although blazars exhibit flaring episodes lasting days to weeks, these short-duration events become completely smeared out over the 3.75 Gyr travel time to Earth. Small variations in boost factor during a flare correspond to tiny velocity differences ($\sim$$10^{-9}$) for the boosted dark matter particles. Over the long journey, these small velocity differences accumulate into arrival time spreads of years. Consequently, particles from a flare that originally lasted only days or weeks arrive at Earth dispersed over many years, effectively erasing the original time structure. Therefore, we observe only the time-averaged luminosity of the blazar, justifying our steady-state approximation.
% In [1]: import numpy as np
% In [3]: gamma = lambda v: 1./np.sqrt(1-v**2)
% In [11]: gamma(v*(1+1e-9))/gamma(v)
% Out[11]: 1.0000500030051336
%\KM{This should emphasize why this matters-- as I understand it it is that the long distance means that even tiny differences in velocity will smear out time signatures of e.g. flares}
Taking these considerations into account, our analysis can include the full range of boosted DM particles without practical limitations from travel-time considerations.

%*****************************************************************
\subsection{\label{sec:modelling}Blazar jet physics}
%*****************************************************************
%Lepto-hadronic modeling; Jet parameters (bulk Lorentz factor, viewing angle, etc.); Proton spectrum and luminosity; Particle acceleration mechanisms

The jet modeling in this paper relies on multiple parameters, each with varying degrees of certainty. Below, we systematically examine these parameters to assess their reliability and impact on our results.

The redshift of the source, and consequently its distance, has been firmly determined by Ref.~\cite{1802.01939}.

The mass of the central black hole (BH) in \txsname~ is estimated to be $M_\mathrm{BH} \approx 6.3\times10^{8} M_{\odot}$, where $M_{\odot}$ denotes the solar mass. This estimate is based on the assumption that the host galaxy is a typical giant elliptical with an absolute R magnitude of $M(R) \approx -22.9$ \cite{Padovani_2022}. Consequently, this corresponds to an Eddington power $L_{\mathrm{Edd}} \approx 7.9 \times 10^{46}$ erg s$^{-1}$.

The motion of the relativistic particles in the blazar jet can be approximately described according to the \virg{blob geometry}~\cite{Dermer:2009zz} in which the accelerated particles are confined within spherical regions (blobs) that move with speed $v_{\Baryon} = \beta_{\Baryon} c$, and bulk Lorentz factor $\Gamma_{\Baryon} = (1-\beta_{\Baryon}^2)^{-1/2}$ along the jet axis. The latter is, in general, inclined at an angle $\theta_\text{l.o.s.}$ with respect to our line-of-sight (l.o.s.). Within each  blob, the individual particles move isotropically relative to the blob center of mass frame.
 
In our analysis, we adopt the leptohadronic model used in~\cite{TXS, TXS_2} to describe the jet emission of blazar \txsname. This model characterizes the electromagnetic energy spectrum, i.e., the energy of the photons emitted by the accelerated charged particles (primarily electrons and protons) in the blazar's jet. The model assumes a homogeneous and isotropic proton energy spectrum, following a single power-law with slope $\alpha_p$ within the energy range $\gamma'_{\min, p} \leq E_p'/m_p \leq \gamma'_{\max, p}$, where $E'_p$ is the proton energy in the blob's rest frame. By fitting the observed electromagnetic spectral energy distribution, Ref.~\cite{TXS,TXS_2} determines the bulk Lorentz factor $\Gamma_{\rm B}$, the spectral index $\alpha_p$, the minimum and maximum Lorentz factors $\gamma'_{\min, p}$ and $\gamma'_{\max, p}$, and the proton luminosity $L_p$ that normalizes the energy power-law spectrum.

The angle $\theta_\text{l.o.s.}$ for blazars is generally known to be small, with $\theta_\text{l.o.s.} < 15 -20^{\circ}$ (e.g. Ref.~\cite{UP_1995,AGNReview}). However, precise values for individual objects are uncertain and difficult to derive. In the analysis of \cite{TXS, TXS_2}, it is assumed that the bulk Lorentz factor of the jet $\Gamma_{\rm B}$ equals $\mathcal{D}/2$, where the latter is the Doppler factor $\mathcal{D} \equiv [\Gamma_{\rm B}\left(1-\beta_{\rm B}\cos\theta_\text{l.o.s.}\right)]^{-1}$. 
This implies $\theta_\text{l.o.s.} = 0^{\circ}$, which is highly improbable and would preclude superluminal motion in this source \cite{Padovani_2022_2}. This contradicts observations, as superluminal motion with $\beta_{\rm app} = 1.07\pm0.14$ has been measured \cite{Lister_2021}. 
A further complication is that the jet parameters are very likely frequency-dependent, with $\Gamma_{\rm B}$ and $\mathcal{D}$ being much larger at high energies compared to the radio band (e.g. \cite{Padovani_2022_2}).

%We then assume $\mathcal{D} = 40$, which is the
%mean value for this parameter derived by the lepto-hadronic spectral energy 
%distribution modelling done by \cite{TXS, TXS_2}, and use two values of 
%$\theta_\text{l.o.s.}$ equal to $5^{\circ}$ and $10^{\circ}$, which are very 
%blazar-like. $\Gamma_{\rm B}$ is then derived from the definition 
%$\mathcal{D} = [\Gamma_{\rm B}\left(1-\beta_{\rm B}\cos\theta_\text{l.o.s.}\right)]^{-1}$,
%where $\beta_{\rm B} = \sqrt{1-\Gamma_{\rm B}^{-2}}$ is the jet speed. TBD
We therefore assume $\mathcal{D} = 35$, which is the smallest value for this parameter 
derived by the lepto-hadronic spectral energy distribution modeling of \cite{TXS_2},
as this implies the largest possible $\theta_\text{l.o.s.} = \arcsin(1/\mathcal{D}) \sim 
1.64^{\circ}$ (see Appendix A of \cite{UP_1995}). For this choice, and 
based on the definition $\mathcal{D}$, 
%, where $\beta_{\rm B} = (1-\Gamma_{\rm B}^{-2})^{1/2}$ is the jet speed, %Already defined
it can be shown that $\Gamma_{\rm B} = \mathcal{D}$. This means that the 
lowest value for the proton power in Table~1 of Ref.~\cite{TXS_2}, which has been derived
assuming $\Gamma_{\rm B} = \mathcal{D}/2$, needs to be multiplied by a factor of 4 for
consistency, giving $L_p = 6.4\times 10^{48}\,\text{erg}/\text{s}$. We take the other blazar jet parameters at the best-fit values given in \cite{TXS_2}. The entire list of parameters and corresponding values considered in our work are summarized in Table~\ref{Tab:HadronicModel}. 

With the parameters in the table, we estimate the number of protons ejected per time, kinetic energy $T_p$ and solid angle $\Omega_p$, in the rest frame of an observer that sees the blobs moving with velocity $\beta_{\Baryon}$ along the jet axis, as \cite{Wang:2021jic}
\begin{equation}\label{eq:CRSpectrum}
\begin{split}
    \frac{d\Gamma_p}{dT_pd\Omega_p} 
    =&\, \frac{c_p}{4\pi}\,\left(1+\frac{T_p}{m_pc^2}\right)^{-\alpha_p}\\ &\times\frac{\beta_p(1-\beta_p\beta_{\Baryon}  \mu)^{-\alpha_p} \Gamma_{\Baryon}^{-\alpha_p}}{\sqrt{(1-\beta_p \beta_{\Baryon} \mu)^2 - (1-\beta_p^2)(1-\beta_{\Baryon}^2)}},
    \end{split}
\end{equation}
where $m_p \simeq 0.938\,\text{GeV}/c^2$ is the proton mass, $\beta_p =[1 - m^2_p/(T_p + m_p )^2]^{1/2}$ is the proton speed in units of $c$, $\mu$ is the cosine of the angle of motion with respect to the jet axis, and $c_p$ is a normalization constant, which we fix by requiring the matching with the fitted proton luminosity $L_p$~\cite{Granelli:2022ysi}. \\
%\AG{We are neglecting the effects of magnetic fields on the trajectory of the particles in the jet, assuming, in practice, a ballistic motion. Shall we comment on this?}\\

\setlength{\tabcolsep}{19pt}
\newcolumntype{C}{@{}>{\centering\arraybackslash}X}
\begin{table}
\centering
\begin{tabularx}{\linewidth}{|@{}C|C@{}|}
    \hline
    \rowcolor[gray]{.95}
    \multicolumn{2}{|@{}c@{}|}{\bf Lepto-Hadronic Model Parameters}\\
    \hline
    \hline
     \rule{0pt}{2.5ex} 
    ~~~~~~~Parameter (unit) & \txsname\\
    \hline
    \rule{0pt}{2.5ex} 
     ~~~~~~~$z$   &0.3365 \\
     ~~~~~~~$d_L$ (Mpc)    & $\sim 1750$\\
     ~~~~~~~$M_{\BH}$ ($M_\odot$) 
     &$6.3\times10^{8} $ \\
%          $t_{\BH}$ (yr) 
%     &$\sim 10^{9} $ \\
          ~~~~~~~$\mathcal{D}$  & 35\\
     ~~~~~~~$\theta_\text{l.o.s.}\, (^\circ)$  & $1.64$  \\
     ~~~~~~~$\Gamma_{\rm B}$ & 35\\  
     %$\beta_{\rm B}$ & 0.9996 \\
%     $\beta_{\rm B}$ & $\sqrt{1-\Gamma_{\rm B}^{-2}}$  \\
%     $\mathcal{D}$  & $[\Gamma_{\rm B}\left(1-\beta_{\rm B}\cos\theta_\text{l.o.s.}\right)]^{-1}$\\
     ~~~~~~~$L_p$ (erg/s)  & $6.4\times 10^{48}$\\
     %$L_e$ (erg/s)  & $? \times 10^{44^\star}$ \\
    ~~~~~~~$\alpha_p$   & $2.0$\\
     %\alpha_e$   & $2.0$\\
    ~~~~~~~$\gamma'_{\min,\,p}$ & 1.0 \\
     ~~~~~~~$\gamma'_{\max,\,p}$ & $5.5\times10^{7^\star}$\\
     %$\gamma'_{\min,\,e}$ & 500\\
     %$\gamma'_{\max,\,e}$ & $1.3\times10^{4^\star}$\\
     ~~~~~~~$c_p$ ($\text{s}^{-1}\text{sr}^{-1}\text{GeV}^{-1}$)  & $2.08 \times 10^{47}$\\
     %$c_e$ ($\text{s}^{-1}\text{sr}^{-1}\text{GeV}^{-1}$)  & $2.42 \times 10^{50}$\\
    \hline
    \end{tabularx}
\caption{{\bf Parameters currently used in the BBDM code for \txsname~\cite{TXS, TXS_2}.} These parameters were partially derived from~\cite{Wang:2021jic, Granelli:2022ysi}.
The quantities flagged with a star ($^\star$) correspond to mean values computed from the ranges given in the second column of Table 1 of Ref.~\cite{TXS_2}.
%In the model fitting of \cite{TXS, TXS_2}, the assumption of $\mathcal{D} 
%= 2\Gamma_{\Baryon}$ is used for TXS 0506+056. 
The resulting values of the normalization constants $c_{e,\,p}$, as well as the redshift $z$ \cite{Keivani:2018rnh}, luminosity distance $d_L$ for $H_0 = 70$, $\Omega_{\rm m}  = 0.3$, and $\Omega_{\rm \Lambda} = 0.7$, BH mass $M_{\BH}$ \cite{Padovani_2022}.}
%and related DM halo mass $M_{\DM}$ for the considered source are also reported.}
%\red{Which cosmological model has been used? For Ho = 70, $\Omega_{\rm m}  = 0.3$, and $\Omega_{\rm \Lambda} = 0.7$ I get 1773.7 Mpc}. %SOME VALUES STILL TO BE UPDATED. 
%\AG{I have updated the values of $c_p$ and the luminosity distance, and commented out the parameters related to the electrons which we do not need. In Ref.~\cite{Keivani:2018rnh} they give $d_L\approx 1750\,\text{Mpc}$ assuming a \virg{consensus cosmology}.}}
\label{Tab:HadronicModel}
%\end{center}
\end{table}

%*****************************************************************
\subsection{\label{sec:DM_profile}Dark matter density profile}
%*****************************************************************
%Three-zone model description
%Profile parameters and assumptions
%Comparison with other profile models

A necessary ingredient to compute the DM flux boosted by blazar jets is the column density integral
\begin{equation}
    \Sigma_{\text{DM}}(r) = \int_{r_\min}^{r} \! \mathrm{d}r'\, \rho_{\text{DM}}(r')
\end{equation}
where $\rho_{\DM}$ is the DM density around the blazar, $r$ is the radial distance from the center of the blazar, and $r_\text{min}$ is the minimum distance along the jet direction considered in the integration. In order to estimate $\Sigma_{\DM}(r)$ we need to make assumptions about the DM profile around the blazar. 

Since the DM limits we derive depend heavily on the DM density profile, particularly in the inner regions near the black hole, our final results are strongly sensitive to assumptions about the DM halo density structure. In this section we examine three DM profile models: the Navarro-Frenk-White (NFW) profile, the Gondolo-Silk profile, and what we call the ``three zone model,'' which we will use to compute our final DM limits.

Figure \ref{fig:profile} shows the dark matter density and integrated column density for the three models. 

\subsubsection{Navarro-FrenkWhite} 
The NFW profile~\cite{Navarro:1996gj} is often used as the default model for galactic DM halos. It takes the form:
\begin{equation}
\label{eq:NFW}
\rho_{\text{NFW}}(r) = \rho_s \frac{r_s/r}{(1 + r/r_s)^2},
\end{equation}
where \( \rho_s \) is the characteristic density (a normalization parameter that sets the density scale) and \( r_s \) is the characteristic scale radius (which determines the transition point between the inner and outer regions of the density profile). 
The characteristic density $\rho_s$ is defined~\cite{Gentile:2007sb} by requiring that the total mass within the virial radius $r_\text{vir}$ equals the virial mass:
\begin{equation}
\label{eq:Mvir}
M_\text{vir} = \delta_\text{vir} \frac{4}{3} \pi r_\text{vir}^3 \rho_c,
\end{equation}
where \( \delta_\text{vir} \simeq 100 \) is the virial overdensity in the standard cosmological model~\cite{Bryan:1997dn}, and \( \rho_c = 1.053672 \times 10^{-5} \, h^{-2} \, (\text{GeV}/c^2)\,\text{cm}^{-3} \) is the critical density of the Universe. Here, \( h \simeq 0.674 \) is the Hubble expansion rate in units of 100~km/s/Mpc~\cite{PDG_Workman:2022ynf}.

Equating the volume integral of $\rho_{\text{NFW}}(r)$ with Eq.~\eqref{eq:Mvir} yields:
\begin{equation}
\label{eq:rho_s}
\rho_s = \frac{\delta_\text{vir}}{3} \, \frac{(r_\text{vir}/r_s)^3}{\ln(1 + r_\text{vir}/r_s) - r_\text{vir}/(r_s + r_\text{vir})} \, \rho_c.
\end{equation}
The characteristic radius $r_s$ and the virial radius $r_\text{vir}$ entering Eq.~
\eqref{eq:rho_s} are expressed in terms of the virial mass as\cite{Wechsler:2005gb, Klypin:2010qw}
\begin{eqnarray}
\label{eq:rs}
    r_s &\simeq& 25.4\,\text{kpc}\left(\frac{M_\text{vir}}{10^{12}M_\odot}\right)^{0.46},\\
    \label{eq:cvir}
    r_\text{vir} &\simeq& 157\,\text{kpc}\left(\frac{M_\text{vir}}{10^{12}M_\odot}\right)^{1/3}.
\end{eqnarray}

The virial mass of the DM halo is estimated via the following relation which stems from numerical simulations \cite{DiMatteo:2003zx} (see also \cite{Ferrarese:2002ct, Baes:2003rt}): %\AG{Are there any more recent Refs.?}: 
\begin{equation}
M_{\text{vir}} \simeq 10^{12}M_\odot\left(\frac{M_{\BH}}{7\times10^{7}M_\odot}\right)^{3/4},
\end{equation}
giving $M_{\text{vir}} \simeq 5.2\times10^{12} M_\odot$ for the parameters in Table~\ref{Tab:HadronicModel}. Using this value in Eqs.~\eqref{eq:cvir} and \eqref{eq:rs}, we get $r_\text{vir}\simeq 273\,\text{kpc}$ and $r_s \simeq 54 \,\text{kpc}$, while from Eq.~\eqref{eq:rho_s},  $\rho_s \simeq 0.10 \,(\text{GeV}/c^2)\,\text{cm}^{-3}$.

The strength of the NFW profile lies in its close agreement with observed rotation velocity curves at large radii. However, at small radii, the DM profile is poorly constrained, making the NFW profile an extrapolation of large-scale trends to small scales without strong empirical support. Furthermore, numerical simulations indicate that the NFW profile is not expected to hold at small radii, since baryonic processes such as stellar feedback and gravitational interactions with the central stellar component tend to modify the inner density profile~\cite{Gnedin2004Aug,Merritt2007Feb,Shapiro2022Aug}.

\subsubsection{Gondolo-Silk}
Below the scale of tens of parsecs, DM is expected to form dense spikes. In their pivotal study~\cite{Gondolo:1999ef}, Gondolo and Silk (GS) showed that, assuming adiabatic accretion of the BH with DM particles gravitating in circular orbits around the central BH, an initial DM density profile of the form $\rho_{\DM} \propto r^{-\gamma}$ is modified into the steeper profile
\begin{equation}\label{eq:rho_prime}
   \rho_\text{spike}(r)  \propto  \left(\frac{R_\text{spike}}{r}\right)^\alpha,
\end{equation}
where $\alpha = (9-2\gamma)/(4-\gamma)$ and $R_\text{spike}$ is the radial extension of the spike. For $\gamma = 1$ as in the NFW profile, the GS spike has slope $\alpha_{\GS} = 7/3$. The existence of spikes around super massive BHs has recently received observational support from Ref.~\cite{chan2024first}, which found evidence for a GS spike at low radii and NFW behavior at large radii for the supermassive BH in blazar OJ 287.

However, numerical simulations generally predict that the spike is modified when additional physical processes are included, such as baryonic effects, noncircular orbits, and displacement of the BH from the halo center~\cite{Ullio2001}. Stellar scattering effects on the dark matter density spike have been extensively studied using the Fokker-Planck equation~\cite{Gnedin2004Aug}, two-fluid approximations, and $N$-body simulations~\cite{Shapiro2022Aug, Merritt2007Feb}. 

These different methodologies consistently show that gravitational scattering of DM by the stellar component becomes important within the radius of influence of the black hole. This radius, $R_\text{infl}$, defines the region where the black hole's gravity dominates the gravitational pull of other objects, like stars in a galaxy, and marks the scale within which stellar scattering introduces an additional baryonic suppression factor $\rho_\text{B}(r) \propto (R_\text{infl}/r)^{\alpha_{\text{Baryon}}}$ with $\alpha_{\text{Baryon}} = 3/2$ that multiplies the original GS spike profile.
$R_\text{infl}$ is given in terms of the stellar velocity dispersion of the galaxy's bulge $\sigma_\star$:
\begin{equation}
    R_\text{infl} = \frac{G M_{\BH}}{\sigma_\star^2},
\end{equation}
where $G$ is Newton's gravitational constant. For the stellar velocity dispersion we have the following relation in terms of $M_{\BH}$~\cite{Kormendy:2013dxa}:
\begin{equation}
    \sigma_\star \simeq 154.5\,\text{km}\,\text{s}^{-1}\left(\frac{M_{\BH}}{10^8M_\odot}\right)^{0.23},
\end{equation}
which implies that for our value of $M_{\BH}$ (Table~\ref{Tab:HadronicModel}) $R_\text{infl} \simeq 49\,\text{pc}$.

A further modification to the spike arises from annihilation of DM particles, which can deplete the spike during the lifetime of the BH, $t_{\BH}$. Self-annihilation effects are primarily confined to the inner regions of dark matter halos due to the density-squared dependence of the annihilation rate. Self-annihilation imposes a limiting density, $\rho_\text{core}$, above which dark matter is depleted faster than it can accumulate, given by
\begin{equation}
    \label{eq:rho_core}
    \rho_\text{core} \simeq 
    \frac{m_\chi}{\langle\sigma_{\chi\chi} v_\text{rel} \rangle \,t_{\BH}},
\end{equation}
where $m_\chi$ is the mass of the DM particle and $\langle \sigma_{\chi\chi} v_\text{rel} \rangle$ is the thermally averaged annihilation cross section times relative velocity. For this analysis, we assume self-annihilation effects are negligible within the radius of influence, effectively taking $\langle \sigma_{\chi\chi} v_\text{rel} \rangle \to 0$. This assumption maximizes the dark matter density available for boosting but should be reconsidered for models with significant self-interaction cross sections. %As a first benchmark situation, we select the case for either very young black holes or non-annihilating dark matter, so that $\rho_{\rm core} \rightarrow \infty$. Alternatively, we consider as a benchmark the value $\rho_{\rm core} \simeq 10^{9}\,\text{GeV}\,\text{cm}^{-3}$ which can be obtained for, e.g., $m_\chi=1\,\text{GeV}$, $\langle \sigma_{\chi\chi}v_\text{rel}\rangle = 3\times 10^{-26}\text{cm}^{3}s^{-1}$, as for a thermal DM relic, and $t_{\BH} = 10^9\,\text{yr}$.

\subsubsection{3-Zone Model}
The DM profile used in this work, $\rho_{\text{3ZM}}(r)$, implements a 3-zone profile similar to that used in Ref.~\cite{Cline:2023tkp}. It includes baryonic effects up to $R_\text{infl}$, a GS spike up to $R_\text{spike}$, and transitions to NFW at larger radii. The full profile is given as:

\begin{widetext}
\begin{equation}
     \rho_{\text{3ZM}}(r) =
     \left(1-\frac{4R_{S}}{r}\right)^{3} \rho_{\text{NFW}}(R_\text{spike}) \,\times
     \begin{cases}
     0 & r \leq 4R_S;\\
     \left(R_\text{spike}/R_\text{infl}\right)^{\alpha_{\text{GS}}}\left( R_\text{infl}/r\right)^{\alpha_{\text{Baryon}}} &  4R_S < r \leq R_\text{infl};\\
    \left(R_\text{spike}/r\right)^{\alpha_{\text{GS}}}  & R_\text{infl} < r \leq R_\text{spike}; \\
     \rho_{\text{NFW}}(r)/\rho_{\text{NFW}}(R_\text{spike}) & r > R_\text{spike},
     \end{cases}
\end{equation}
\end{widetext}

\noindent where $4R_S$ is the radius within which DM is captured by the BH, with $R_S = 2G_N M_{\text{BH}}/c^2$ being the Schwarzschild radius. The normalization factors ensure continuity at $R_\text{infl}$ and $R_\text{spike}$. The volume integral of the 3-zone DM profile is dominated by the NFW contribution 
\begin{equation}
    \int_{4R_S}^{r_\text{vir}} \! \mathrm{d}r\, r^2 \rho_{\text{3ZM}}(r) \simeq \int_{0}^{r_\text{vir}} \! \mathrm{d}r\, r^2 \rho_{\text{NFW}}(r) = \frac{M_\text{vir}}{4\pi}
\end{equation} 
ensuring that numerical simulations are not significantly affected by the presence of spikes in the inner regions.

The spike extension $R_\text{spike}$ is determined by requiring that the DM mass enclosed within the radius of influence equals the BH mass, corresponding to the typical uncertainty in BH mass measurements in this region.

Annihilation effects can be incorporated using the limiting density $\rho_\text{core}$ as
 \begin{equation}
     \rho_{\text{DM}}(r) = \frac{\rho_{\text{3ZM}}(r)\rho_{\text{core}}}{\rho_{\text{3ZM}}(r) + \rho_{\text{core}}}.
 \end{equation}

In the absence of DM annihilations, the column density integral is dominated by the contributions at low radii, including in the GS and baryonic regions, and approaches
\begin{equation}
    \Sigma_{\text{DM}} \simeq \left(\frac{R_\text{spike}}{R_\text{infl}}\right)^{\alpha_{\text{GS}}-1}\left(\frac{R_\text{infl}}{r_\min}\right)^{\alpha_{\text{Baryon}}-1}\frac{r_s\rho_s}{\alpha_{\text{Baryon}}-1}.
\end{equation}
With self-annihilating DM, the column density is instead dominated by the core contribution $\Sigma_{\text{DM}} \simeq \rho_\text{core}R_\star$, where $R_\star$ is the distance at which $\rho_{\text{3ZM}}(R_\star) = \rho_\text{core}$.

\begin{figure}
    \centering
    \includegraphics[width=1\columnwidth]{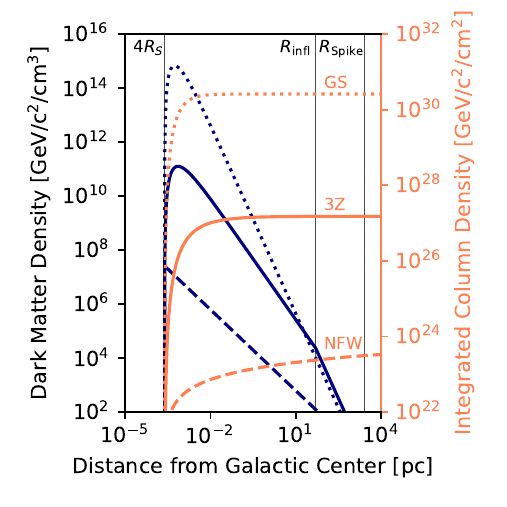}
    \caption{Dark matter density (blue lines, left $y$-axis) and integrated column density (orange lines, right $y$-axis) as a function of distance from the galactic center for three different dark matter profiles. Dashed lines refer to the NFW DM profile model, dotted lines to GS one, and solid lines show the 3-zone model. Vertical thin lines mark the capture radius ($4R_{\text S}$), radius of influence ($R_{\text{infl}}$), and spike extension ($R_{\text{spike}}$), defining the boundaries between the three zones.}
    \label{fig:profile}
\end{figure}

%********************************
\section{\label{sec:source_to_detector}From source to Earth}
%********************************

\subsection{\label{sec:dm_p_scattering}BBDM flux}

To calculate the flux of blazar-boosted dark matter reaching Earth, we need to model the scattering process between DM particles and protons in the blazar jet. 
The same approach applies to DM-nucleus interactions at the detector. 
Throughout this work, for baseline demonstration, we follow the result-reporting conventions of dark matter direct detection experiments~\cite{Baxter:2021pqo}, modeling the DM-proton scattering with a fixed total cross section that is energy-independent.
For DM-proton scattering, we use the dipole form factor
\begin{equation}
    G_\text{D}(Q^2)=\frac{1}{(1+Q^2/\Lambda^2)^2},
\end{equation}
where $Q^2$ is the squared momentum transfer and $\Lambda \simeq 0.84\,\text{GeV}$ for the proton. While ideally we would use a form factor derived from the weak charge distribution, accurate measurements of this would require precise neutrino-proton scattering data, which is not currently available. In our kinematic regime the momentum transfer is much smaller than $\Lambda$, so $G_D(Q^2)\approx 1$ and the dipole form factor introduces no relevant energy dependence.

For interactions with heavy nuclei in xenon detectors, we use the standard Helm form factor, which accounts for the spatial distribution of nucleons within the nucleus and is the established choice for xenon-based dark matter experiments~\cite{Lewin:1995rx,Baxter:2021pqo}.

If the cross section $\sigma_{\chi p}$ of the elastic scattering between the DM particle $\chi$ and the proton is assumed to be isotropic in the center-of-mass (c.m.) rest frame and independent of the c.m.~energy, the BBDM flux $\Phi_\chi$ for a dark matter particle with mass $m_\chi$ is given by \cite{Wang:2021jic, Granelli:2022ysi}:
\begin{equation}\label{eq:spectrumDM}
    \frac{d\Phi_\chi}{dT_\chi} =\frac{\Sigma^\tot_{\DM}\sigma_{\chi p}}{m_\chi d_L^2}\int_0^{2\pi} d\phi_s \int_{T_p^\min(T_\chi)}^{T_p^\max}\frac{dT_p}{T_\chi^\max(T_p)}\frac{d\Gamma_p}{dT_p d\Omega_p},
\end{equation}
where $\Sigma_{\DM}^\text{tot} = \Sigma_{\DM}(r_\max)$ is the total dark matter column density along the line of sight through the blazar jet, $r_\max$ is the maximal radial extension of the blazar jet, $d_L$ is the luminosity distance, $\phi_s$ is the azimuthal angle of scattering with respect to the l.o.s., $T_p^\min(T_\chi)$ is the minimal kinetic energy required to have an outgoing DM particle with kinetic energy $T_\chi$, $T_p^\max$ is the maximal energy of the protons available in the blazar jet, and $T_\chi^\max (T_p)$ is the maximal kinetic energy that the DM has after a forward scattering with a proton of energy $T_p$. The proton spectrum needs to be evaluated at the proper angle that leads to scattering of DM toward the Earth. Consequently, since, in our case, $\theta_\text{l.o.s.}\neq 0$, the proton spectrum inherits a nontrivial dependence on $\phi_s$ and $T_p$ (see, e.g., Ref.~\cite{Granelli:2022ysi} for more details). 

Using the attenuation and detector response described in the following sections, we examined how the high energy part of the BBDM spectrum contributes to the 
predicted BBDM rate at the detector. We compared the recoil spectrum obtained by integrating the flux to arbitrarily high energies with the spectrum obtained when truncating it at $T_\chi \simeq 1$~GeV. Since the difference is negligible, we truncated the BBDM flux to include energies only up to $T_\chi = 1$~GeV. This can be understood intuitively: because the DM mass is many orders of magnitude smaller than the xenon nuclear mass, even relativistic DM can transfer only a small fraction of its kinetic energy in an elastic collision. As a result, the maximum recoil energy saturates at a few 
hundred keV, and the very high energy part of the BBDM flux does not produce additional detectable events in liquid xenon. This conclusion is driven entirely by recoil kinematics and the steeply falling BBDM flux, and therefore remains valid even if the DM–nucleus cross section has a mild energy dependence.

\subsection{\label{sec:attenuation}Earth attenuation}

For dark matter particles with relatively large interaction cross sections, Earth's crust significantly attenuates the boosted DM flux before it reaches underground detectors. This effect can substantially impact detector sensitivity and must be accounted for.

We model this attenuation using the analytical energy-loss method from Eq. (11) of Ref.\cite{Bringmann:2018cvk}. While more detailed Monte Carlo simulations using \textsc{Darkprop} \cite{Xia:2021vbz} are available, we use the analytical approach due to computational constraints. However, the choice of attenuation method is not the leading source of uncertainty in our limits; the dark matter density profile has a much greater impact on the final constraints.

\begin{figure}
    \centering
    \includegraphics[width=\columnwidth]{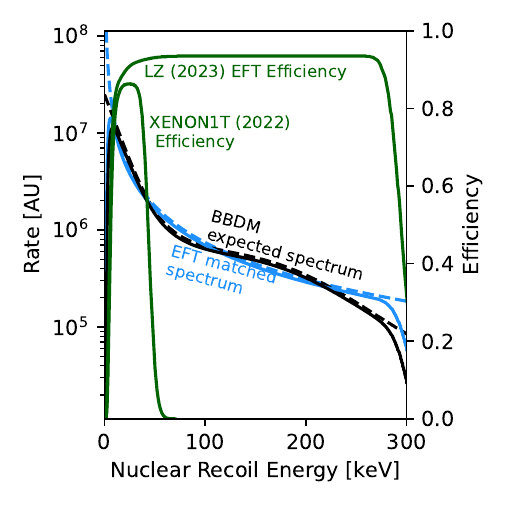}
    \caption{Comparison of nuclear recoil energy spectra in liquid xenon for a \SI{46.4}{\kevm} WIMP boosted by \txsname. In black is the expected energy deposition spectrum in liquid xenon, while in blue is the signal from the LZ EFT analysis that best fits the blazar-boosted energy deposition, a \SI{1}{\tevm} WIMP coupling to nuclear spin ($\mathcal{O}_4$). Green lines show the efficiency for the XENON1T WIMP search reaching $\sim$\SI{50}{\keV}~\cite{XENON:2018voc}, and for the LZ EFT analysis where the energy range is extended to above $\SI{250}{\keV}$~\cite{LZ:2023lvz}. The comparison illustrates how blazar-boosted dark matter can produce signals in energy ranges that extend beyond typical WIMP search windows.
}
    \label{fig:spectrum}
\end{figure}

%********************************
\section{\label{sec:detector}At the detector}
%********************************
The expected signal in the underground DM detector for a given DM spectrum relies on the DM-nucleon interaction and the detector response to that recoil. 
The DM-nucleon interaction is the same spin-independent interaction of the blazar jet acceleration as described above. The expected signal from a recoil is encoded in the detector response, which combines the microphysics model of how many scintillation and ionization quanta are expected from an elastic recoil with a model of the detector. The probability for a recoil to be recorded, reconstructed and pass analysis quality cuts is reported by the experiments as the efficiency, illustrated in Fig.~\ref{fig:spectrum}, and may be used to compute the expected signal rate in the detector for a certain DM flux and interaction. For the full interpretation of their data, experiments rely on full detector models of the expected signal in their scintillation and ionization observables, which enables them to also take the varying shape of DM recoil models into account.

%********************************
\section{\label{sec:methods_limits}Statistical interpretation using public data}
%********************************

Data from liquid xenon time-projection-chambers (LXe TPCs) are interpreted in terms of specific dark matter-xenon interaction models, most commonly assuming a spin-independent WIMP-nucleon interaction, with a standard halo model velocity distribution~\cite{Baxter:2021pqo} and considering nuclear recoils between \SI{3}{\keV} to $\sim \SI{40}\keV$ as shown in the Fig.~\ref{fig:spectrum} (see the line for XENON1T). 

The simplest method for computing approximate limits is what we refer to as ``rate-matching,'' following the approach in Ref.~\cite{Wang:2021jic}. This method finds the cross section required for the blazar-boosted signal rate in the detector to match the published WIMP limit event rate. This approach takes \sout{almost} no spectral information into account other than detector efficiency. In this paper, we improve accuracy and sensitivity by incorporating as much spectral information as possible into the search.

We use XENON1T and LZ public data to derive constraints on the DM–nucleon cross section.

\begin{figure}
    \centering
    \includegraphics[width=1\columnwidth]{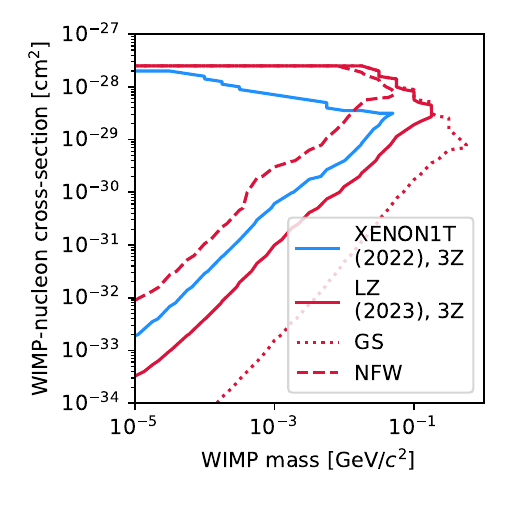}
    \caption{Constraints on the DM cross section as function of WIMP mass for blazar-boosted dark matter. The blue line shows the constraint using the XENON1T open likelihood~\cite{xenon1tbinference,XENON:2018voc}, while the red lines indicate recasted limits from the LZ EFT search~\cite{LZ:2023lvz}. Full lines illustrate the constraint using the three-zone model DM column density, while the dashed and dotted lines show the constraints for the NFW and GS DM density model, respectively. The upper edge of the constrained region is due to the attenuation of the DM flux by the Earth.
    We show the same limits for different dark-matter density profiles to highlight that the choice of halo model can strongly affect the resulting constraints.
}
    \label{fig:limit}
\end{figure}

{\bf XENON1T case study.} The XENON1T collaboration has provided a complete statistical model of their tonne-year search ~\cite{XENON:2018voc,xenon1tbinference}. We use this model to compute constraints on BBDM (Fig. \ref{fig:limit}) and to validate the ``rate-matching'' approach. We find qualitative agreement between the two methods, with deviations of around a factor of 2 in the rightmost corner of the exclusion region. These larger deviations occur where attenuation most strongly affects the recoil spectra.

{\bf LZ case study.} BBDM may yield significantly higher recoil energies than the typical WIMP signature. The LZ collaboration has developed a nuclear recoil model to extend their search region to almost \SI{300}{keV}~\cite{LZ:2023lvz} to constrain effective field theory (EFT) interactions. Combined with LZ's generally higher sensitivity to WIMPs, this extended energy range significantly improves constraints on boosted DM, such as in Ref.~\cite{LZ:2025iaw}. 
To incorporate spectral information from LZ, we use $\chi^2$ minimization to find the effective field theory (EFT) operator from Ref.~\cite{LZ:2023lvz} that best matches each Earth-attenuated BBDM recoil spectrum. We then find the BBDM cross section that produces the rate excluded by LZ. We perform this matching and compute exclusions across a grid of BBDM masses for the assumed BBDM model and cross sections, with results shown in Fig.~\ref{fig:limit}. We validated this matching approach by comparing it with a more conservative approach where instead of the best-match scale we demand that the BBDM spectrum at the limit exceeds the matched spectrum in a recoil energy region that contains $90\%$ of the signal. 

Since the integrated column density appears as a multiplicative factor in Eq.~\ref{eq:spectrumDM}, the exclusion limits on cross section scale as the square root of the integrated column density in the absence of attenuation effects. In the region where attenuation impacts the spectrum, however, the rate in the detector varies faster with changes in cross section, and conversely, limits on the DM cross section change less with changes in flux. Both of these regions can be seen in Fig.~\ref{fig:limit}, where constraints from the same experiments are shown for the three DM column densities shown in Fig.~\ref{fig:profile}. 
Below WIMP masses$\sim\SI{3e-4}{\gevm}$, the upper limit scales as expected in the no-attenuation case. 
It is worth highlighting that, in the low-mass range, the sensitivity improves because the dark-matter number density increases, leading to a higher boosted flux from the blazar jet. 
In standard WIMP searches, by contrast, the sensitivity drops at low masses since halo dark matter moves nonrelativistically and produces recoil energies below the detector threshold. 
In the blazar-boosted scenario, however, the jet acceleration can provide relativistic dark-matter particles, enabling detectable nuclear recoils even for sub-MeV masses.

\textbf{About the constant cross section.}
In setting exclusion limits we adopted a constant–cross section approximation (as in Ref.~\cite{Bringmann:2018cvk}). In reality, the full relativistic DM–nucleon scattering cross section retains energy dependence~\cite{Dent:2019krz, de2025boosted}---even in the heavy-mediator regime. This behavior is explicitly shown in Refs.~\cite{ema2021neutrino,bell2024cosmic}, where mediator masses of order a GeV still lead to clear departures from an energy–independent interaction.

Using a constant $\sigma_{\chi N}$ therefore slightly underestimates Earth attenuation in the large–cross–section regime, shifting the attenuation ``ceiling’’ toward mildly larger (more optimistic) values. This happens because an energy–dependent cross section increases with DM kinetic energy, leading to more scattering in the crust and hence stronger attenuation than the constant–$\sigma$ assumption predicts. Although physically meaningful, this region of parameter 
space is already excluded~\cite{aalbers2025new}, so this shift does not affect the interpretation of the BBDM limits derived here.

At small cross sections---where Earth attenuation is negligible---the constant cross–section approximation is conservative: any realistic energy dependence would increase the predicted scattering rate and therefore strengthen (not weaken) 
the limits. In practice, these theoretical uncertainties are much smaller than the astrophysical uncertainties in the dark–matter density profile near the black hole, which dominate the variation in the final constraints.

%********************************
\section{\label{sec:conclusion} Conclusions} 

%********************************
We present the first blazar-boosted dark matter search incorporating full detector response modeling, using public data from XENON1T and LZ to constrain light dark matter accelerated by \txsname. We place model-dependent exclusion regions on the dark-matter–nucleon cross section for$\sim$\SI{1}{\mevm} dark matter between \SI{5.8e-31}{\cm\squared} and \SI{6.3e-29}{\cm\squared} using XENON1T data, and between \SI{9.9e-32}{\cm\squared} and \SI{2.5e-28}{\cm\squared} from LZ EFT searches. Our analysis demonstrates that relativistic jets in active galactic nuclei provide an additional mechanism for probing sub-GeV dark matter that would otherwise be inaccessible to liquid xenon detectors.

Our key finding is that astrophysical uncertainties, particularly the dark-matter density profile near the supermassive black hole, represent the primary limitation on sensitivity rather than detector systematics. The constraints vary by more than an order of magnitude depending on the adopted density profile, highlighting the critical importance of inner halo structure modeling. We demonstrate how to recast direct detection results using full spectral information rather than simple rate-matching, and find that extending the nuclear recoil energy range, as demonstrated by LZ's coverage up to $\sim$\SI{300}{\keV} compared to XENON1T's $\sim$\SI{50}{\keV} cutoff, provides substantial improvements in sensitivity since boosted dark matter deposits significantly higher energies than typical halo WIMPs.
We also compared the 
full–detector–response method with a simple rate-matching approach and found that 
the two generally agree at the qualitative level, differing noticeably only in 
regions where attenuation becomes important.

These results establish blazar-boosted dark matter as a complementary probe for sub-GeV dark matter, extending sensitivity to mass ranges inaccessible to conventional halo searches. However, blazar-boosted searches are fundamentally limited by astrophysical uncertainties for now, whereas conventional galactic halo searches are primarily constrained by detector sensitivity. 
Our work motivates both improved astrophysical modeling of galactic centers and extension to multiple blazar sources. We encourage direct detection experiments to publish their full results in formats that facilitate recasting for alternative dark matter scenarios. The framework developed here can be readily applied to next-generation detectors such as DARWIN, and the strong coupling between dark matter searches and galactic structure studies opens new avenues for multimessenger constraints on both dark matter physics and the central regions of massive galaxies.

%********************************
\section*{Acknowledgments}
%********************************
We thank Alessandro Granelli for providing the blazar-boosted dark matter flux code used in this work and for helpful discussions at different stages of the project.
We thank Andrea Macciò and Matteo Nori for helpful discussions about the dark matter density profile toward the center of the blazar, Benjamin Davis for guidance on extending spiral galaxy scaling relations to elliptical galaxies, and Maria Petropoulou for insights into blazar jet physics.
The work of Y.X. has received funding from the European Union’s Horizon Europe research and innovation programme under the Marie Skłodowska-Curie grant agreement No. 101126636.

\section*{Data availability}
The data that support the findings of this article are openly available~\cite{xenon1tbinference, LZ:2023lvz}.

% \begin{enumerate}
%     \item Using New model gives worse results by like half an order of magnitude
% \end{enumerate}
% Halo Changes
% \begin{enumerate}
%     \item toward a less simplistic density profile model $ > 3$-zone model (more widely accepted by a larger community)
%     \item Still tons of uncertainty and heavily profile dependent and especially interior profile dependent
% \end{enumerate}

% Leptohadronic changes
% \begin{enumerate}
%     \item Changes in Doppler factor and proton luminosity
% \end{enumerate}

% What else should a future analysis do?
% \begin{enumerate}   
%     \item Leptohadronic model of other blazar with corresponding neutrino event detection.
%     \item What is the discussed DARWIN prospected sensitivity? Potential for discovery.
% \end{enumerate}

\bibliographystyle{apsrev4-1}
\bibliography{main}

\end{document}